\documentclass[prd, twocolumn, nofootinbib, floatfix]{revtex4-1}

\usepackage{amsmath}
\usepackage{graphicx}
\usepackage{dcolumn}
\usepackage{bm}
\usepackage{epsfig}
\usepackage{amssymb,latexsym,mathrsfs}
\usepackage{graphicx}
\usepackage{color}
\usepackage{hyperref}

\hypersetup{
    colorlinks=true,
    linkcolor=red,
    citecolor=blue,
} 

\newcommand{\be}{\begin{equation}}
\newcommand{\ee}{\end{equation}}
\newcommand{\bs}{\begin{split}} 
\newcommand{\bea}{\begin{eqnarray}}
\newcommand{\eea}{\end{eqnarray}}

\newcommand{\gm}{G_{\rm matter}} 
 
\newcommand{\gl}{G_{\rm light}}

\newcommand{\fs}{f\sigma_8}

\newcommand{\alm}{\alpha_M} 
\newcommand{\alb}{\alpha_B}

\newcommand{\ms}{M_\star^2}

\begin{document}

\title{Limited Modified Gravity} 

\author{Eric V.\ Linder${}^{1,2}$} 
\affiliation{
${}^1$Berkeley Center for Cosmological Physics \& Berkeley Lab, 
University of California, Berkeley, CA 94720, USA\\ 
${}^2$Energetic Cosmos Laboratory, Nazarbayev University, 
Nur-Sultan, Kazakhstan 010000}

\begin{abstract} 
We systematically assess several limiting cases of modified gravity, 
where particular theoretical or observational conditions hold. This 
framework includes the well known scalar-tensor gravity and No Slip 
Gravity and No Run Gravity, and we extend it to three new limits: Only 
Run, Only Light, and Only Growth Gravities. These limits give simplifications 
that allow deeper understanding of modified gravity, including demonstration 
that gravitational effects on light and matter can have opposite signs 
in their deviation from general relativity. We also show observational 
predictions for the different cosmic  structure growth rates $\fs$ and the ratio of 
gravitational wave standard siren luminosity to photon standard candle 
luminosity distance relations, defining a new statistic $D_G$ that emphasizes 
their complementarity and ability to distinguish models. 
\end{abstract}

\date{\today} 

\maketitle

\section{Introduction}

Modification of general relativity is a viable explanation for current 
cosmic acceleration and has several further predictable consequences 
beyond the expansion history, such as the change in large scale structure 
growth relative to the expansion history, change in light deflection 
behavior, gravitational slip (distinction between time-time and space-space 
metric gravitational potentials), and altered propagation of gravitational 
waves relative to light (even for the same speed of propagation). 

This offers a rich array of signatures that current and upcoming observations 
can test (see recent reviews such as 
\cite{1806.10122,1809.08735,1902.10503,1907.03150}). 
However, unlike the expansion history, which for a broad variety  
of models can be described by a few parameters (e.g.\ the matter density, 
dark energy equation of state today $w_0$, and dark energy time variation 
$w_a$ have been shown to reconstruct the expansion history to 0.1\% 
accuracy \cite{calde}), modified gravity theories need not just a few 
parameters but four functions of time in general, in addition to the 
expansion history \cite{gubitosi,eft1,glpv,bellsaw,eft2}. 

Still, if our first goal is to detect any modifications of gravity, one  
can parametrize the effects on the observables. For example, binning the 
modified Poisson equation gravitational strength $\gm$ in redshift 
delivers subpercent accuracy for growth of structure with just 2--3 
parameters \cite{misha1,misha2}. Comparison of observational data in 
expansion vs in growth offers another useful avenue for alerts, e.g.\ 
\cite{huterer1709.01091,conjoin}. 

A middle path is also desirable, where one has some closer connection 
to theory, and a way of seeing covariances between the several observable 
effects mentioned in the opening paragraph. One would like to narrow 
down the multiple functional freedom but still have a viable theory that 
can predict the multiple effects simultaneously. We call this limited modified 
gravity. 

Section~\ref{sec:limit} introduces limited modified gravity in two forms, 
from the theory side and the phenomenology side, giving a systematic 
summary of the limiting cases. In Sec.~\ref{sec:quant} we investigate 
the main effects of the three new cases on gravitational strengths, 
sound speed stability, etc. Section~\ref{sec:results} then propagates  
this to observables, showing how these models can be tested, and introducing 
the $D_G$ statistic, revealing 
signatures beyond general relativity. We conclude in Sec.~\ref{sec:concl}.

\section{The System of Limited Modified Gravity} \label{sec:limit} 

Many individual theories of modified gravity exist, but under certain 
physical assumptions the most general scalar-tensor theory involving 
no more than second derivatives is the Horndeski class, e.g.\ 
\cite{horn,deffayet,gubitosi,glpv}. This will contain  
four functions of the scalar field $\phi$ and its kinetic term $X$, 
$G_i(\phi,X)$ for $i=2$--5. No general principle for how to specify the 
functional forms is known. Within linear theory one can equivalently 
phrase the gravitational action in effective field theory terms, which 
can in turn be treated by property functions, which are functions of time, 
e.g.\ \cite{bellsaw,eft2}. No general principle for how to specify those 
functional forms is known. 

The situation improves somewhat if one applies the constraint that 
gravitational waves must propagate at the speed of light, as strongly 
suggested by observations \cite{gwspeed1,gwspeed2,gwspeed3}. 
This (in the simplest 
interpretation) removes $G_5$ and makes $G_4(\phi)$, independent of $X$. 
On the property function side, it removes $\alpha_T$. It is still difficult 
to see how to choose, or parametrize, the Horndeski functions, but there 
are some interesting perspectives on the property function side. Some 
modified gravity theories have specific relations between the property 
functions $\alpha_i(t)$, for example $f(R)$ gravity has $\alb(t)=-\alm(t)$. 

For the theory avenue, therefore, we work with the property functions. 
Since the kineticity $\alpha_K$ generally has negligible observational 
impact \cite{bellsaw,nscmb}, we mostly ignore it (setting it to $10^{-4}$, 
see \cite{nscmb}), 
leaving two functions: 
the Planck mass running $\alm$ and the braiding $\alb$. We explore limited 
modified gravity through the limits where one or the other of these 
functions is zero (when both are zero then general relativity holds), or 
one is determined by the other. That is, we focus on how one function 
determines the observational signatures. 

For the phenomenology avenue, we work with the modified Poisson equations 
that relate the metric gravitational potentials to the observable 
large scale structure spatial density perturbations: 
\bea 
\nabla^2\psi&=&4\pi G_N\delta\rho\times \gm\\ 
\nabla^2(\psi+\phi)&=&8\pi G_N\delta\rho\times\gl \ . 
\eea 
The first equation governs the growth of structure, with a 
gravitational strength $\gm$ relative to Newton's constant $G_N$, 
and the second governs the deflection 
of light, with a gravitational strength $\gl$. Both $\gm$ and $\gl$ 
are functions of time. We explore limited modified gravity here through 
the limits where one or the other of these functions is unity (when both  
are unity then general relativity holds), or they are equal to each other. 

Other quantities that come from the two main functions are the 
gravitational slip, 
\be 
\bar\eta=\frac{\gm}{\gl}=\frac{2\psi}{\psi+\phi}\ , 
\ee 
measuring the offset between the metric potentials or gravitational 
strengths, and the sound speed of scalar fluctuations, $c_s$, 
important for testing stability of the theory, with $c_s^2\ge0$ required. 

The relations between the phenomenological approach and the theory 
quantities are (with $\alpha_T=0$): 
\bea 
\gm&=&\frac{m_p^2}{M_\star^2}\frac{(2+2\alm)(\alb+2\alm)+2\alb'}{(2-\alb)(\alb+2\alm)+2\alb'} \label{eq:gmgen} \\ 
\gl&=&\frac{m_p^2}{M_\star^2}\frac{(2+\alm)(\alb+2\alm)+2\alb'}{(2-\alb)(\alb+\
2\alm)+2\alb'} \label{eq:glgen} \\ 
\bar\eta&=&\frac{(2+2\alm)(\alb+2\alm)+2\alb'}{(2+\alm)(\alb+2\alm)+2\alb'} 
\label{eq:etagen} \\ 
c_s^2&=&\frac{1}{\alpha_{K}+3\alpha_{B}^{2}/2} \left[\left(1-\frac{\alpha_B}{2}\right)\left(\alb+2\alpha_{M}\right)\right. \label{eq:csgen} \\
    &\qquad&\left.+\frac{(H\alpha_B)'}{H}+\frac{\rho_m}{H^2}\left(1-\frac{m_p^2}{\ms}\right)+\frac{{\rho}_{\rm de}(1+w)}{H^2}\right] \ , \notag
\eea 
where $m_p$ is the usual (constant) Planck mass, $M_\star(t)$ is the 
running Planck mass, prime denotes $d/d\ln a$ with $a$ the scale factor, 
$H$ is the Hubble parameter, $\rho_m$ is the matter density, $\rho_{\rm de}$ 
the effective dark energy density, and $w$ the dark energy equation of state. 

We can now identify several special limiting cases that will hopefully 
offer some interesting insights into how each limit affects observations, 
and provide benchmark theories that are tractable to test. 

From the 
theory side, these are when $\alm=0$ and when $\alb=0$. The first of these 
is known as No Run Gravity, since the Planck mass is constant, and was 
investigated in \cite{nrg}. The second has not been specifically studied, 
and is explored here as 
``Only Run Gravity''. There are also well known theories relating nonzero 
$\alm$ and $\alb$, such as various scalar-tensor theories like $f(R)$ 
gravity, Brans-Dicke, or chameleon theories where $\alb=-\alm$, and 
No Slip Gravity where $\alb=-2\alm$ \cite{nsg,nscmb}. One can write more  
generally $\alb=R\alm$ where $R$ is constant \cite{island}, but this 
does not seem to have clear physical significance unlike the previous 
two cases. No Slip Gravity, as the name suggests, has the special property 
that the gravitational slip $\bar\eta=1$, as in general relativity, despite 
modifying other parts of gravitation. The $\alb=-\alm$ scalar-tensor 
cases reduce the effect on light deflection to arise only from the Planck 
mass change. Thus, three of the four limited modified gravity classes 
of this sort are known, and we will explore the fourth. 

From the phenomenology side, we can either set $\gl=m_p^2/M_\star^2$, 
i.e.\ the only effect on light is from the Planck mass, or do the same 
for $\gm=m_p^2/M_\star^2$. The former leads to either the above scalar-tensor 
theories or to No Slip Gravity. The latter leads to No Slip Gravity. 
If we set $\gm=\gl$ then the only solutions are either general relativity 
(where they are both unity) or No Slip Gravity (where they are both 
$m_p^2/M_\star^2$). These properties demonstrate that No Slip Gravity 
is a particularly physically meaningful limit (even apart from having 
no slip). 
If we force $\gm$ or $\gl$ to be completely unaffected, i.e.\ to be unity, 
then we have two new theories, which we call Only Light Gravity and Only 
Growth Gravity, since only one of $\gl$ or $\gm$ is modified. Of course if 
both are unity then we have general relativity. Imposing such conditions 
on $\gm$ or $\gl$ defines a relation between $\alm$ and $\alb$ as we 
discuss below. 

Table~\ref{tab:models} summarizes the limited modified gravity classes, 
and gives the expressions for $\gm$, $\gl$, $\bar\eta$, $\alm$, and $\alb$ 
in each one.

\begin{table*}[tb]
    \centering
\begin{tabular}{|l|c|c|c|c|c|c|}
\hline 
Model & $\gm$ & $\gl$ & $\bar\eta$ & $\alpha_M$ & $\alpha_B$ & Reference \\ 
\hline 
\rule{0pt}{1.1\normalbaselineskip}Scalar-tensor & $\gm$ & $\frac{m_p^2}{M_\star^2}$ & $\bar\eta$ & $\alm$ & 
$-\alm$ & e.g.~\cite{0610532,0805.1726,1002.4928,1901.08690} \\  
\rule{0pt}{1.1\normalbaselineskip}No Slip & $\frac{m_p^2}{M_\star^2}$ & $\frac{m_p^2}{M_\star^2}$ & 1 & 
$\alm$ & $-2\alm$ & \cite{nsg} \\  
\rule{0pt}{1.1\normalbaselineskip}No Run  & $\gm$ & $\gm$ & 1 & 0 & $\alb$ & \cite{nrg} \\ 
\rule{0pt}{1.1\normalbaselineskip}Only Run & $\frac{m_p^2}{M_\star^2}(1+\alm)$ & $\frac{m_p^2}{M_\star^2} 
\left(1+\frac{\alm}{2}\right)$ & $\frac{1+\alm}{1+\alm/2}$ & $\alm$ & 0 & 
new \\ 
\rule{0pt}{1.1\normalbaselineskip}Only Light & 1 & $\gl$ & $1/\gl$ & $\alm$ & dif.eq.($\alm$) & new \\ 
\rule{0pt}{1.1\normalbaselineskip}Only Growth & $\gm$ & 1 & $\gm$ & $\alm$ & dif.eq.($\alm$) & new \\ 
\hline 
\end{tabular} \\  
\caption{Several limited modified gravity models are of special interest, 
having specific characteristics in either theoretical or observational  
functions. The lower three are new and analyzed in this article. The 
notation ``dif.eq.($\alm$)'' denotes that $\alb$ is determined from $\alm$ 
by a differential equation. 
}
\label{tab:models} 
\end{table*}

\section{Gravitational Functions} \label{sec:quant} 

We focus in the rest of the article on the three new theories, and here 
relate the key quantities in each.

\subsection{Only Run Gravity} \label{sec:onlyrun} 

When $\alb=0$ then the scalar sector is not braided with the tensor  
sector and the effective dark energy does not cluster on subhorizon  
scales (see \cite{bellsaw} for details), however there is still modification 
to matter perturbation growth. The main quantities are determined by $\alm$ 
(and the expansion history), with 
\bea 
\gm&=&\frac{m_p^2}{\ms}(1+\alm)\\ 
\gl&=&\frac{m_p^2}{\ms}\left(1+\frac{\alm}{2}\right)\\ 
\bar\eta&=&\frac{1+\alm}{1+\alm/2}\ . 
\eea 
Note that $\alb=0$ implies in the Horndeski approach that 
$G_{4\phi}=XG_{3X}$. This can be compared to No Slip Gravity which 
has $G_{4\phi}=-XG_{3X}$. 

In the early universe we wish all quantities to restore to general 
relativity, to preserve primordial nucleosynthesis and the cosmic 
microwave background (CMB) results, so we want $\alm(a\ll1)\to0$. 
At late times, if we seek a de Sitter state then all time variations 
must stop, so $\ms$ freezes and again $\alm(a\gg1)\to0$, since 
$\alm\equiv d\ln\ms/d\ln a$. This tells us that at early times 
$\gm\to\gl\to1$, $\bar\eta\to1$ and at late times 
$\gm\to\gl\to m_p^2/\ms$, $\bar\eta\to1$. Moreover, $\bar\eta$ will have 
its maximum deviation from unity when $\alm$ does. Thus a deviation 
of $\alm$ (and hence growth and other effects) from general relativity in 
the observable epoch will also give slip in that epoch. 

One can also see from Eq.~(\ref{eq:csgen}) that the sound speed squared 
will be proportional to $\alm$ at early times, requiring $\alm(a\ll1)\ge0$ 
for stability. The same holds in the approach to a de Sitter state. Thus 
the simplest model will be of a hill form, as in \cite{nsg}. We adopt 
that here: 
\be 
\alm=\frac{4c_M\,(a/a_t)^\tau}{[(a/a_t)^\tau+1]^2} \ , \label{eq:hill} 
\ee 
which also gives an analytic expression 
\be 
\frac{\ms}{m_p^2}=e^{(2c_M/\tau)(1+\tanh[(\tau/2)\ln(a/a_t)])}\ . 
\ee 
The Planck mass squared therefore goes from 1 in the past to $e^{4c_M/\tau}$ 
in the future. Note the maximum of $\alm$ is $c_M$, occurring at $a=a_t$, 
and $\tau$ is a measure of the transition width. 

Unlike in No Slip Gravity and No Run Gravity, stability at early times 
only depends on the sign of $c_M$, requiring it to be positive, and not 
on the value of $\tau$. We also assume $\alpha_K>0$. What is especially  
interesting is the influence of $\tau$ on $\gm$ and $\gl$. At early times 
$\alm\to 4c_M(a/a_t)^\tau$ and  
\bea 
\gm&\to&1+(\tau-1)\alm \label{eq:gm1}\\ 
\gl&\to&1+(\tau/2-1)\alm\ . \label{eq:gl1} 
\eea 
This implies that for $1<\tau<2$, the gravitational strength deviations 
from general relativity $\gm-1$ and $\gl-1$ will have opposite signs, 
one being weaker and one being stronger than Einstein gravity. This 
gives a direct proof that such a condition is possible, as argued by 
\cite{1607.03113} regarding the conjecture requiring same signs 
\cite{1606.05339}. 

\begin{figure}[tbp!]
\includegraphics[width=\columnwidth]{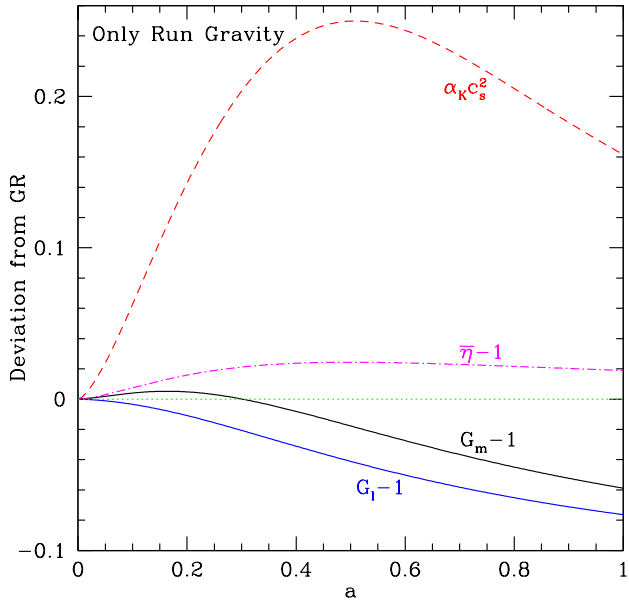}
\caption{
The deviations from general relativity (GR) for the gravitational coupling 
strength for matter $\gm-1$, for light $\gl-1$, 
the gravitational slip $\bar\eta-1$, and the sound speed squared $c_s^2$ 
for Only Run Gravity are plotted vs scale factor. Here we take the hill 
form Eq.~(\ref{eq:hill}) for $\alm$ with $c_M=0.05$, $a_t=0.5$, $\tau=1.5$. 
} 
\label{fig:quant4} 
\end{figure}

Figure~\ref{fig:quant4} exhibits the evolution of the key quantities 
$\gm$, $\gl$, $\bar\eta$, and $c_s^2$. Indeed we see that $\gm-1$ and 
$\gl-1$ can have opposite signs -- gravity for growth is strengthened 
while gravity for light deflection is weakened -- at early times. 
Gravitational slip is present and $c_s^2>0$ shows the theory is stable. 
Extending the numerical evolution further to the future than shown 
verifies that all quantities freeze to constant values (with $\bar\eta\to1$). 

\begin{figure}[tbp!]
\includegraphics[width=\columnwidth]{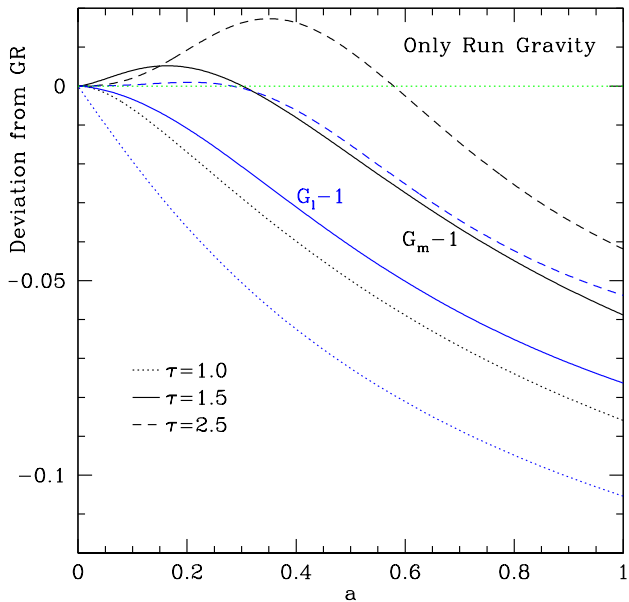}
\caption{
The gravitational strengths $\gm$ and $\gl$ for Only Run Gravity can exhibit 
opposite deviations from general relativity (opposite signs in $\gm-1$ and 
$\gl-1$) for some of their evolution. We show cases for three different 
values of $\tau$, with the behavior following the analytic predictions of 
Eqs.~(\ref{eq:gm1})--(\ref{eq:gl1}). 
} 
\label{fig:quant2tau} 
\end{figure}

Figure~\ref{fig:quant2tau} explores the opposite sign conditions on 
$\gm-1$ and $\gl-1$ in further detail, verifying the analytic derivation 
that this occurs at early times when $1<\tau<2$. Note however that the 
opposite signs can occur more generally at later times, including during 
the key observational window $a\approx 0.25$--0.6.

\subsection{Only Growth Gravity} \label{sec:onlygrow} 

When we limit the modification of gravity to the 
growth sector, leaving light deflection unchanged from general relativity, 
$\gl=1$, this imposes a condition relating $\alb$ to $\alm$ through a 
differential equation, 
\be 
\alb'=(\alb+2\alm)\left[-1+\frac{\alb+\mu\alm}{2(1-\mu)}\right]\ , 
\ee 
where $\mu=m_p^2/\ms$. 

We can plug this back into Eq.~(\ref{eq:gmgen}) to obtain 
\be
\gm=\frac{\alb+\alm(2-\mu)}{\alb+\alm}\ .
\ee
Note however that one must solve the differential equation to obtain
$\alb(\alm)$. 
The early universe limit is $\gm\to1$ so $\mu\to1$, $\alm\to0$, $\alb\to0$.
The de Sitter limit is $\gm\to1$ with 
$\alb\to 2(1-m_p^2/M^2_{\star,{\rm dS}})$ since $\alb'\to0$. 
The stability condition $c_s^2\ge0$ at early times requires $\alm>0$. 
The differential equation is straightforward to solve and we present the 
numerical results below, also checking stability for all times. The 
results are insensitive to initial conditions, as long as 
$\alpha_{B,i}<-\alpha_{M,i}$ (otherwise $\gm$ will diverge when 
$\alb+\alm$ crosses zero). 

\begin{figure}[tbp!]
\includegraphics[width=\columnwidth]{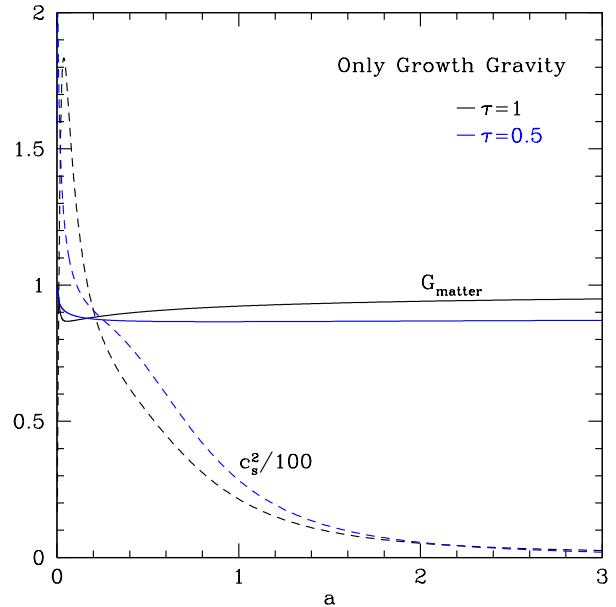}
\caption{
The gravitational coupling strength for matter $\gm$ in Only Growth Gravity 
can be weaker than in general relativity. The evolution $\gm(a)$ is shown 
with the solid curves, and the sound speed squared $c_s^2(a)$ by the 
dashed curves. For each, two values of $\tau$ are exhibited for $\alm$ in 
the hill form, with $c_M=0.03$, $a_t=0.5$. 
} 
\label{fig:qmatter} 
\end{figure}

Figure~\ref{fig:qmatter} shows $\gm$ and $c_s^2$ for the hill form of 
$\alm$ with $c_M=0.03$, $a_t=0.5$, and two different values of $\tau$. 
(Recall that for this model $\bar\eta=\gm$.) 
Note that the modified Poisson equation for growth shows weaker gravity  
than general relativity (while the Poisson equation for light deflection 
is the same as general relativity by construction). The minimum strength 
$G_{\rm matter,min}\approx 1-4c_M$ and then it slowly approaches $\gm\to1$ 
in the de Sitter limit (verified by extending the integration to $a\gg1$); 
also $c_s^2\to0$ in that limit. Only Growth Gravity, like No Slip Gravity 
but unlike many scalar-tensor theories, suppresses growth -- this can 
ease some tensions in $f\sigma_8$ measurements from redshift space 
distortions with respect to $\Lambda$CDM cosmology, and possibly also 
$\sigma_8$ tensions from weak lensing -- making it a theory worth further 
study.

\subsection{Only Light Gravity} \label{sec:onlylight} 

The third new theory is limiting the modification of gravity to the 
light deflection sector, leaving growth unchanged from general relativity, 
$\gm=1$. This again gives a relation for $\alb$ in terms of $\alm$, in  
the form of 
\be  
\alb'=(\alb+2\alm)\left[-1+\frac{\alb+2\mu\alm}{2(1-\mu)}\right]\ . 
\ee 

We can plug this back into Eq.~(\ref{eq:glgen}) to obtain 
\be 
\gl=\frac{\alb+\alm(1+\mu)}{\alb+2\alm}\ .  
\ee 
Again note that one must solve the differential equation to obtain 
$\alb(\alm)$. 
The early universe limit is $\gl\to1$ so $\mu\to1$, $\alm\to0$, $\alb\to0$. 
The de Sitter limit is $\gl\to1$ with 
$\alb\to 2(1-m_p^2/M^2_{\star,{\rm dS}})$, as in the Only Growth Gravity 
case, and again the differential equation is straightforward to solve. 

Only Light Gravity is more difficult, however, in that the denominator 
of $\gl$ involves $\alb+2\alm$ and this is exactly the prefactor in the 
$\alb'$ equation. This means that if at some point in the evolution of 
$\alb$ it reaches or crosses $-2\alm$, as the dynamical equation 
motivates, then the gravitational strength diverges. 
We have not been able to find cases yet where this does not occur 
(e.g.\ trying the hill form for $\alm$, or power law times Gaussians), 
though we also have not found a proof there is no nondivergent solution.

\section{Observational Functions} \label{sec:results} 

These modified gravity theories are highly predictive (in the linear 
regime at least). With the expressions for $\gm$, $\gl$, and  $\ms$ 
one can calculate observables in growth and light propagation. Furthermore, 
\cite{nsg} identified a clear link between predictions for cosmic 
growth and for gravitational wave propagation. Basically, deviations in  
cosmic growth predict deviations in gravitational waves and vice versa. 

This allows an important 
test for modified gravity -- if a signature is seen in growth of large 
scale structure, it could be seen as well in the luminosity distances 
of gravitational wave standard sirens vs standard candles. Such a 
crosscheck is a valuable systematics test; while one might find other 
cosmological model parameters or astrophysical uncertainties that could 
change growth and, say, the CMB or lensing dynamics in a way that mimics 
modified gravity (e.g.\ neutrinos or selection effects), 
such common systematics are much less 
likely with a gravitational wave comparison. 

Therefore in this section we not only look at the observational effects 
on large scale structure growth through the growth rate $\fs$, but also 
their connection to observational effects on gravitational wave propagation. 
Recall that luminosity distances for photon sources, such as supernovae, 
only depend on the background expansion, which we are holding fixed when 
we change gravity from general relativity. However gravitational wave 
propagation is sensitive to the Planck mass running 
\cite{1406.7139,1408.2224,1509.08458,1710.04825,1711.03776,1712.08623,1712.08108,nsg,holz}, and so 
\be 
\frac{d_{L,{\rm GW}}(a)}{d_{L,\gamma}(a)}=\left[\frac{\ms(a=1)}{\ms(a)}\right]^{1/2} \ . 
\ee 

\begin{figure}[tbp!]
\includegraphics[width=\columnwidth]{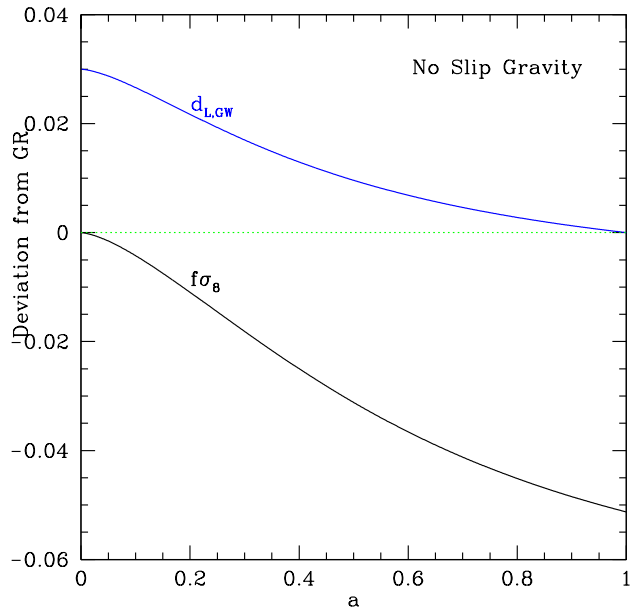}
\caption{
Deviations from general relativity in the cosmic growth and gravitational 
wave distance predictions are connected, and serve as a valuable crosscheck. 
Here the relations are shown for $d_{L,{\rm GW}}^{\rm MG}/d_L^{\rm GR}-1$ 
and $\fs^{\rm MG}/\fs^{\rm GR}-1$ for No Slip Gravity, with model parameters 
$c_M=0.03$, $a_t=0.5$, $\tau=1.5$. Deviations will scale linearly with $c_M$. 
} 
\label{fig:nsggw} 
\end{figure}

Figure~\ref{fig:nsggw} shows the prediction for both probes 
for No Slip Gravity. 
We see the characteristic suppression of growth, at the 3--5\% level, 
relative to general relativity, over the currently measured range of 
redshifts using redshift space distortions as in Fig.~3 of \cite{nsg}. 
But in addition we plot the deviation in luminosity distance to gravitational 
wave standard sirens relative to photon luminosity distances, e.g.\ from 
standardized candles such as Type Ia supernovae. At redshift $z=1$ this 
model predicts a 1\% deviation in $d_L$, concomitant with a 3\% deviation 
in $\fs$. As measurements move to higher redshift, say $z=2$, 
the deviations become 
1.6\% in $d_L$ and 2\% in $\fs$. The numbers given are for $c_M=0.03$ and 
will scale linearly with $c_M$. The key point is that the gravity model 
predicts exactly how they should be related at all redshifts, allowing 
for leverage by combining several low signal to noise measurements.

\begin{figure}[tbp!]
\includegraphics[width=\columnwidth]{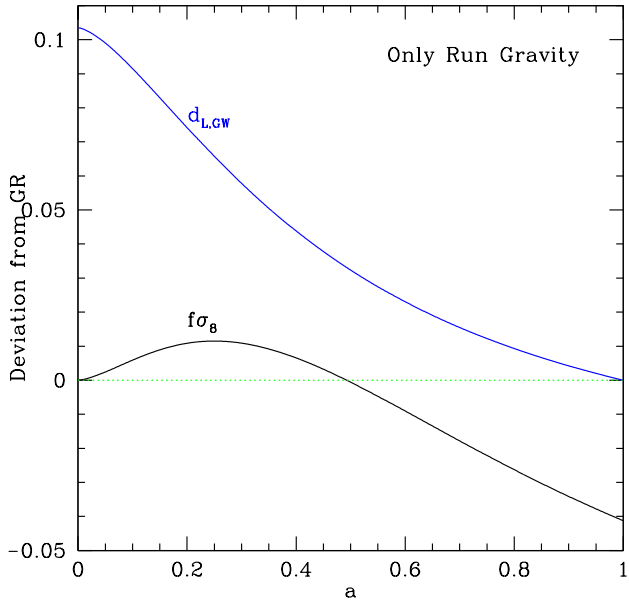}
\caption{
As Fig.~\ref{fig:nsggw} but for Only Run Gravity, with model parameters 
$c_M=0.1$, $a_t=0.5$, $\tau=1.5$. Relatively large values of $c_M$ still 
give viable results for growth, allowing for strong effects on gravitational 
waves. 
} 
\label{fig:onlyrungw} 
\end{figure}

Figure~\ref{fig:onlyrungw} shows the growth and gravitational wave  
quantities for Only Run Gravity. Here, the deviation of the growth 
from general relativity is partially canceled because the gravitational 
strength $\gm$ is enhanced at high redshift, but suppressed at low 
redshift, as seen in Fig.~\ref{fig:quant2tau}. This increases $\fs$ 
relative to general relativity for $a\lesssim0.5$ but decreases it for 
$a\gtrsim0.5$. That allows higher values of Planck mass running amplitude 
$c_M$ to be viable for growth observations. However, the hiding of the 
deviation in growth due to the cancellation does not hold for the 
gravitational wave luminosity distance, which sees simply the enhancement 
of $\ms$ relative to $m_p^2$. Thus the two observational probes work 
extremely well together.

\begin{figure}[tbp!]
\includegraphics[width=\columnwidth]{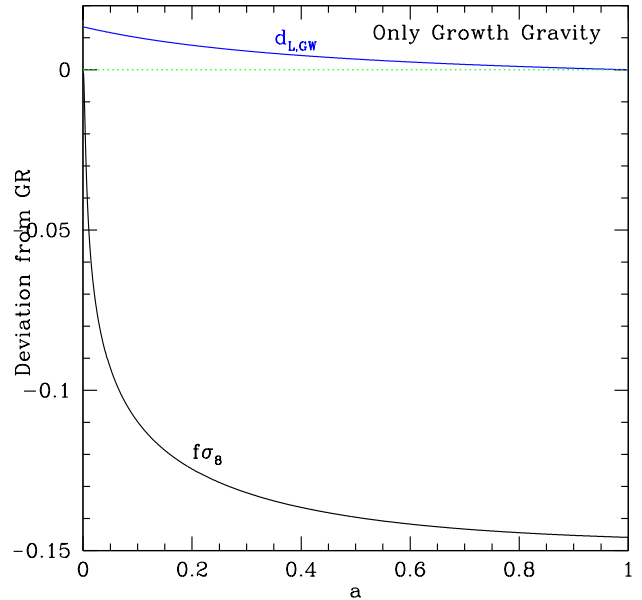}
\caption{
As Fig.~\ref{fig:nsggw} but for Only Growth Gravity, with model parameters 
$c_M=0.01$, $a_t=0.5$, $\tau=1$. Note that the early time, and sustained, 
weakening of $\gm$ as seen in Fig.~\ref{fig:qmatter} have a strong effect 
to suppress growth. This indicates that even small values of $c_M$ can 
have an observable effect on growth, though then the effect on gravitational 
waves becomes negligible. 
} 
\label{fig:onlygrowgw} 
\end{figure}

Figure~\ref{fig:onlygrowgw} shows the growth and gravitational wave  
quantities for Only Growth Gravity. This has a third, distinct behavior 
for the relation between growth and gravitational waves. Due to the 
rapid suppression of $\gm$ at early times, the growth gets off to a slow 
start, and the continued weakness of gravity does not allow it to recover, 
giving a strongly suppressed growth rate in the observational epoch. 
This requires a small value of $c_M$ for viability, which substantially  
reduces the signature of deviation in gravitational waves. However this 
does mean that cosmic growth measurements can probe much smaller $c_M$ 
values than the other models discussed.

Thus we have seen that cosmic growth rate measurements through redshift 
space distortions and gravitational wave luminosity distance measurements 
through standard sirens have great complementarity. The three models we 
discussed in this section have distinct signatures in each, with predictions 
for their respective redshift dependences. Measurements 
through both probes could not only test general relativity but distinguish 
between these classes of gravity models: No Slip Gravity gives discernible 
deviations in each, Only Run Gravity has a larger effect on gravitational 
waves, and Only Growth Gravity has a larger effect on the cosmic growth 
rate. (And of course Only Light Gravity has no effect on growth, only 
on gravitational waves, while No Run Gravity has no effect on gravitational 
waves, but enhances growth.) 

\begin{figure}[tbp!]
\includegraphics[width=\columnwidth]{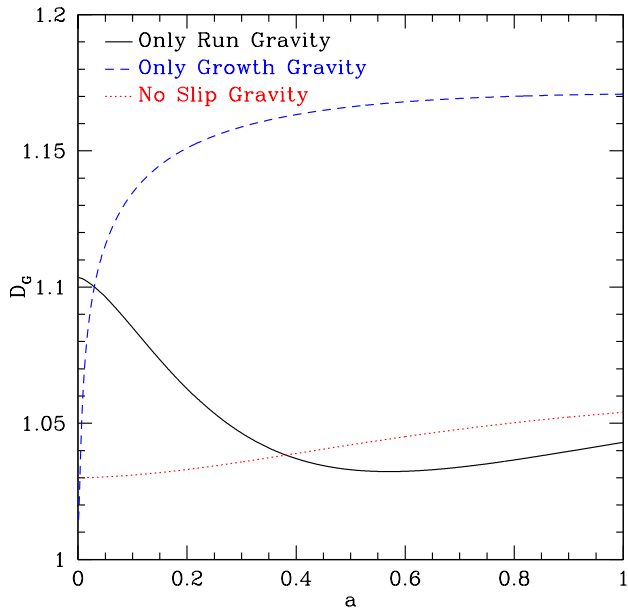}
\caption{
The new $D_G$ statistic, using the complementarity of the gravitational 
wave luminosity distance $d_{L,{\rm GW}}$ and the cosmic matter growth 
rate $\fs$, can clearly distinguish different classes of gravity. Each 
class has a distinct shape in its redshift dependence $D_G(a)$. General 
relativity has constant $D_G=1$. 
} 
\label{fig:dlfsgw} 
\end{figure}

We demonstrate the clear leverage for distinguishing the classes of 
gravity by defining a new statistic, 
\be 
D_G(a)=\frac{d_{L,{\rm GW}}^{\rm MG}/d_L^{\rm GR}}{\fs^{\rm MG}/\fs^{\rm GR}}\ . 
\ee 
In general relativity this is simply a constant with value unity for all 
$a$. However each of the classes of modified gravity we discussed will 
not only show in the $D_G$ statistic deviations from unity (testing general 
relativity), but have a distinct shape with redshift. While scaling $c_M$ 
will change the amplitude, it will not mix the shapes. 
Figure~\ref{fig:dlfsgw} illustrates that indeed the different models are 
highly distinct in the $D_G$ statistic.

\section{Conclusions} \label{sec:concl} 

We assessed in a systematic way limits of modified gravity in terms of 
property functions and observational functions, including introducing 
three new classes of modified gravity. Such limits are simpler than the 
full freedom of gravity theories but are more predictive, and display 
clear signatures that observations can use to test general relativity  
and distinguish between theory classes. 

For the three new theories -- Only Run Gravity, Only Growth Gravity, 
and Only Light Gravity -- we compute the key functions of the 
gravitational strengths for cosmic growth and for light deflection, 
$\gm$ and $\gl$, and the gravitational slip $\bar\eta$ and scalar 
perturbation sound speed squared $c_s^2$. Interestingly, Only  
Run Gravity provides a definite demonstration that the deviations 
from general relativity $\gm-1$ and $\gl-1$ for matter and light 
can have opposite signs, which has been a topic of conjecture. 
These theories can also provide suppressed matter growth, in 
contrast to many scalar-tensor theories and in some accord with 
observations. 

In addition to solving for the evolution of these key functions, 
we also calculate two observational quantities. One is $\fs$, 
the cosmic growth rate for large scale structure perturbations, 
measurable through redshift space distortions in galaxy 
surveys such as DESI \cite{desi}. The other is the luminosity 
distance to gravitational wave standard siren events, 
$d_{L,{\rm GW}}$, which can differ from the photon luminosity  
distance to standard candles such as Type Ia supernovae, 
despite a gravitational wave propagation speed equal to the 
speed of light. 

Conjoined analysis of the two observables, $\fs$ and 
$d_{L,{\rm GW}}$, as introduced by \cite{nsg}, is highly 
insightful. For one thing, they offer a critical crosscheck 
for systematic control. As well, there is a diversity of 
behaviors between the classes of gravity in the magnitude 
of deviations in one vs the other, and predictive power in 
the specific redshift dependence between the two. This 
enables even low signal to noise measurements at individual 
redshifts to combine to give significant evidence to test 
general relativity and distinguish classes of gravity. 
We defined a new statistic $D_G$ to use for the conjoined 
analysis of the two probes, illustrating that it has distinct redshift 
dependence for different classes. 
Future measurements will demonstrate the strong 
complementarity of these probes. 
Other combinations of gravitational wave and large scale structure 
information are discussed in, e.g., \cite{1901.07832,1908.08951}. 

There is still much to understand about modified gravity, 
especially if one starts furthest from the observations with the 
$G_i(\phi,X)$ functions in the Horndeski lagrangian. The relation 
between these functions exhibited by, e.g., Only Run Gravity 
and No Slip Gravity may provide some direction to future 
investigations, but here we focused on quantities closer to the 
observations. The approach of Limited Modified Gravity gives 
a framework that is tractable, predictive, and yet with a range 
of important characteristics that can yield insights when 
confronted with forthcoming data.

\acknowledgments 

This work is supported in part by the Energetic Cosmos Laboratory and by the 
U.S.\ Department of Energy, Office of Science, Office of High Energy Physics, 
under Award DE-SC-0007867 and contract no.\ DE-AC02-05CH11231.



\begin{thebibliography}{99}


\bibitem{1806.10122} 
M. Ishak, Testing General Relativity in Cosmology, Living Rev. Relativ. 22, 
1 (2019) [arXiv:1806.10122] 

\bibitem{1809.08735} 
R. Kase, S. Tsujikawa, Dark energy in Horndeski theories after GW170817: A 
review, Int. J. Mod. Phys. D 28, 1942005 (2019) [arXiv:1809.08735] 

\bibitem{1902.10503} 
P.G. Ferreira, Cosmological Tests of Gravity, Ann. Rev. Astron. Astrophys. 
57, 335 (2019) [arXiv:1902.10503] 

\bibitem{1907.03150} 
N. Frusciante, L. Perenon, Effective Field Theory of Dark Energy: a Review, 
arXiv:1907.03150 

\bibitem{calde} 
R. de Putter, E.V. Linder, Calibrating Dark Energy, JCAP 0810, 042 (2008) 
[arXiv:0808.0189] 

\bibitem{gubitosi} 
G. Gubitosi, F. Piazza, and F. Vernizzi, The effective field theory of 
dark energy, JCAP 1302, 032 (2013) [arXiv:1210.0201] 

\bibitem{eft1} 
J.K. Bloomfield, E.E. Flanagan, M. Park, S. Watson, Dark energy or 
modified gravity? An effective field theory approach, JCAP 1308, 010 
(2013) [arXiv:1211.7054] 

\bibitem{glpv} 
J. Gleyzes, D. Langlois, F. Piazza, F. Vernizzi, Essential building 
blocks of dark energy, JCAP 1308, 025 (2013) [arXiv:1304.4840] 

\bibitem{bellsaw} 
E. Bellini, I. Sawicki, Maximal freedom at minimum cost: linear large-scale 
structure in general modifications of gravity, JCAP 1407, 050 (2014) 
[arXiv:1404.3711] 

\bibitem{eft2} 
E.V. Linder, G. Seng{\"o}r, S. Watson, Is the Effective Field Theory of 
Dark Energy Effective?, JCAP 1605, 053 (2016) [arXiv:1512.06180] 

\bibitem{misha1} 
M. Denissenya, E.V. Linder, Cosmic Growth Signatures of Modified 
Gravitational Strength, JCAP 1706, 030 (2017) [arXiv:1703.00917] 

\bibitem{misha2} 
M. Denissenya, E.V. Linder, Subpercent Accurate Fitting of Modified Gravity 
Growth, JCAP 1711, 052 (2017) [arXiv:1709.08709 ]

\bibitem{huterer1709.01091} 
D. Huterer, D.L. Shafer, Dark energy two decades after: Observables, probes, 
consistency tests, Rep. Prog. Phys. 81, 016901 (2018) [arXiv:1709.01091] 

\bibitem{conjoin}  
E.V. Linder, Cosmic Growth and Expansion Conjoined, Astropart. Phys. 86, 
41 (2017) [arXiv:1610.05321] 

\bibitem{horn} 
G.W. Horndeski, Second-order scalar-tensor field equations in a 
four-dimensional space, Int. J. Theor. Phys. 10 (1974) 363 

\bibitem{deffayet} 
C. Deffayet, X. Gao, D.A. Steer, G. Zahariade, From k-essence to 
generalised Galileons, Phys. Rev. D84 (2011) 064039 [arXiv:1103.3260] 

\bibitem{gwspeed1} 
B. Abbott et al. (LIGO, Virgo Collaborations), 
GW170817: Observation of Gravitational Waves from a Binary Neutron Star 
Inspiral, Phys. Rev. Lett. 119, 161101 (2017) [arXiv:1710.05832] 

\bibitem{gwspeed2} 
B. Abbott et al. (LIGO, Virgo Collaborations), 
Multi-messenger Observations of a Binary Neutron Star Merger, 
ApJ Lett. 848, L12 (2017) [arXiv:1710.05833] 

\bibitem{gwspeed3} 
B. Abbott et al. (LIGO, Virgo Collaborations), 
Gravitational Waves and Gamma-Rays from a Binary Neutron Star Merger: 
GW170817 and GRB 170817A, 
ApJ Lett. 848, L13 (2017) [arXiv:1710.05834] 

\bibitem{nscmb} 
M. Brush, E.V. Linder, and M. Zumalac{\'a}rregui, No Slip CMB, 
JCAP 1901, 029 (2019) [arXiv:1810.12337] 

\bibitem{nrg} 
E.V. Linder, No Run Gravity, JCAP 1907, 034 (2019) [arXiv:1903.02010] 

\bibitem{nsg} 
E.V. Linder, No Slip Gravity, JCAP 1803, 005 (2018) [arXiv:1801.01503] 

\bibitem{island} 
M. Denissenya, E. Linder, Gravity's Islands: Parametrizing Horndeski 
Stability, JCAP 1811, 010 (2018) [arXiv:1808.00013] 

\bibitem{0610532} 
Y-S. Song, W. Hu, I. Sawicki, The Large Scale Structure of f(R) Gravity, 
Phys. Rev. D 75, 044004 (2007) [arXiv:astro-ph/0610532] 

\bibitem{0805.1726} 
T.P. Sotiriou, V. Faraoni, f(R) Theories Of Gravity, Rev. Mod. Phys. 82, 
451 (2010) [arXiv:0805.1726] 

\bibitem{1002.4928} 
A. De Felice, S. Tsujikawa, f(R) theories, Living Rev. Rel. 13, 3 (2010) 
[arXiv:1002.4928] 

\bibitem{1901.08690}
I. Quiros, Selected topics in scalar-tensor theories and beyond, 
Int. J. Mod. Phys. D 28, 1930012 (2019) [arXiv:1901.08690] 

\bibitem{1607.03113} 
E.V. Linder, Challenges in Connecting Modified Gravity Theory and 
Observations, Phys. Rev. D 95, 023518 (2017) [arXiv:1607.03113] 

\bibitem{1606.05339} 
L. Pogosian, A. Silvestri, What can Cosmology tell us about Gravity? 
Constraining Horndeski with Sigma and Mu, Phys. Rev. D 94, 104014 (2016) 
[arXiv:1606.05339] 

\bibitem{1406.7139} 
I.D. Saltas, I. Sawicki, L. Amendola, M. Kunz, 
Anisotropic Stress as a Signature of Nonstandard Propagation of 
Gravitational Waves, Phys. Rev. Lett. 113, 191101 (2014) [arXiv:1406.7139] 

\bibitem{1408.2224} 
V. Pettorino, L. Amendola, Friction in Gravitational Waves: a test for 
early-time modified gravity, Phys. Lett. B 742, 353 (2015) [arXiv:1408.2224] 

\bibitem{1509.08458} 
L. Lombriser, A. Taylor, Breaking a Dark Degeneracy with Gravitational 
Waves, JCAP 1603, 031 (2016) [arXiv:1509.08458]

\bibitem{1710.04825} 
A. Nishizawa, Generalized framework for testing gravity with 
gravitational-wave propagation. I. Formulation, Phys. Rev. D 97, 104037 
(2018) [arXiv:1710.04825] 

\bibitem{1711.03776} 
S. Arai, A. Nishizawa, Generalized framework for testing gravity with 
gravitational-wave propagation. II. Constraints on Horndeski theory, 
Phys. Rev. D 97, 104038 (2018) [arXiv:1711.03776] 

\bibitem{1712.08623} 
L. Amendola, I. Sawicki, M. Kunz, I.D. Saltas, 
Direct detection of gravitational waves can measure the time variation of 
the Planck mass, JCAP 1808, 030 (2018) [arXiv:1712.08623] 

\bibitem{1712.08108} 
E. Belgacem, Y. Dirian, S. Foffa, M. Maggiore, 
Gravitational-wave luminosity distance in modified gravity theories, 
Phys. Rev. D 97, 104066 (2018) [arXiv:1712.08108] 

\bibitem{holz} 
M. Lagos, M. Fishbach, P. Landry, D.E. Holz, 
Standard sirens with a running Planck mass, Phys. Rev. D 99, 083504 
(2019) [arXiv:1901.03321] 

\bibitem{desi} 
DESI Collaboration, The DESI Experiment Part I: Science,Targeting, and 
Survey Design, arXiv:1611.00036 

\bibitem{1901.07832} 
L. Amendola, Y. Dirian, H. Nersisyan, S. Park, 
Observational Constraints in Nonlocal Gravity: the Deser-Woodard Case, 
JCAP 1903, 045 (2019) [arXiv:1901.07832] 

\bibitem{1908.08951} 
S. Mukherjee, B.D. Wandelt, J. Silk, 
Probing the theory of gravity with gravitational lensing of gravitational 
waves and galaxy surveys, 
arXiv:1908.08951 


\end{thebibliography}
\end{document}